# Probing the limits of superconductivity

D. R. Strachan, M. C. Sullivan, and C. J. Lobb
University of Maryland, Department of Physics, Center for Superconductivity Research, College Park, MD 20742-4111

**ABSTRACT**

DC voltage versus current measurements of superconductors in a magnetic field are widely interpreted to imply that a phase transition occurs into a state of zero resistance. We show that the widely-used scaling function approach has a problem: Good data collapse occurs for a wide range of critical exponents and temperatures. This strongly suggests that agreement with scaling alone does not prove the existence of the phase transition. We discuss a criterion to determine if the scaling analysis is valid, and find that all of the data in the literature that we have analyzed fail to meet this criterion. Our data on YBCO films, and other data that we have analyzed, are more consistent with the occurrence of small but non-zero resistance at low temperature.

## 1. INTRODUCTION

The most striking property of superconductors is perfect lossless conductivity below a critical temperature $T_c$. Since the discovery of superconductivity, much research has been done to understand the limits of perfect conductivity.[1] In type-I superconductors, it was found that there are no losses only below a critical current density $J_c$, a critical field $H_c$, and only (strictly) at zero frequency; see Fig. 1(a). The picture in conventional type-II superconductors, Fig. 1(b), was the same as in type-I superconductors below a lower critical field $H_{c1}$. (For $H<H_{c1}$ in a type-II superconductor, or $H<H_c$ in a type-I superconductor, magnetic field is excluded from the sample; this is the Meissner effect.) For the field range $H_{c1}<H<H_{c2}$, it was generally believed that, while the resistivity might be immeasurably small, it was never strictly zero. Non-vanishing resistivity was believed to result from thermally-activated flux motion, which would generate voltages, leading to dissipation.

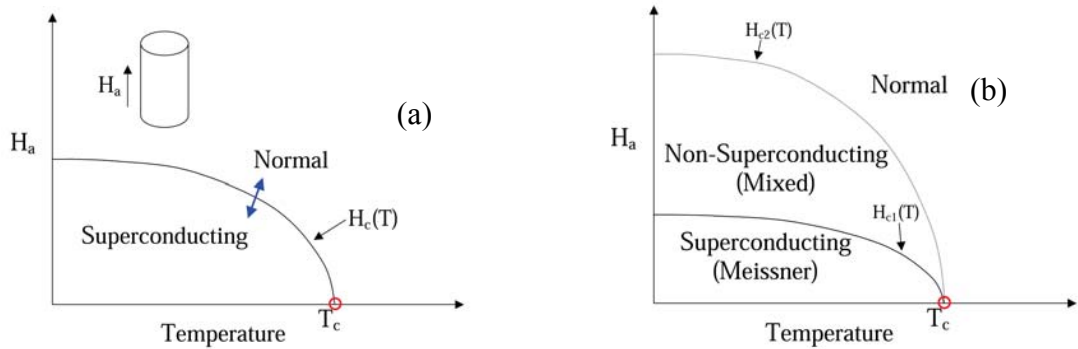

Fig. 1: Phase diagrams for type I (a) and type II superconductors (b). $H_a$ is the applied magnetic field.

The discovery of the high-temperature superconductors provided a new opportunity to explore the limits of superconductivity. Because their transition temperatures are an order of magnitude higher than conventional superconductors, thermal fluctuations play a much more visible role in the superconducting to normal transition,[2] broadening the temperature range over which the transition can be observed. In addition, the high-$T_c$ superconductors have higher normal-state resistivities than conventional superconductors. Voltage signals are expected to scale with the normal-state resistivity even in the transition region, again making the superconducting to normal transition easier to study.

Starting with the pioneering theoretical work of Fisher[3] and experimental work of Koch *et al.*,[3] a new picture of the transition for type-II superconductors has emerged.[5-10] Contrary to the understanding based on conventional superconductors, it is now generally believed that a transition to a true zero-resistivity state occurs in the presence of a magnetic field.[1,3,11,12] Various theories have been proposed for this phase transition, including a vortex glass transition,[3,11] which is predicted to occur when disorder in the superconductor is uncorrelated, and a Bose glass transition,[12] which is predicted to occur in the presence of correlated disorder. While these theories apply to different situations, both predict that the resistivity should be zero below a transition temperature. A schematic phase diagram is shown in Fig. 2.

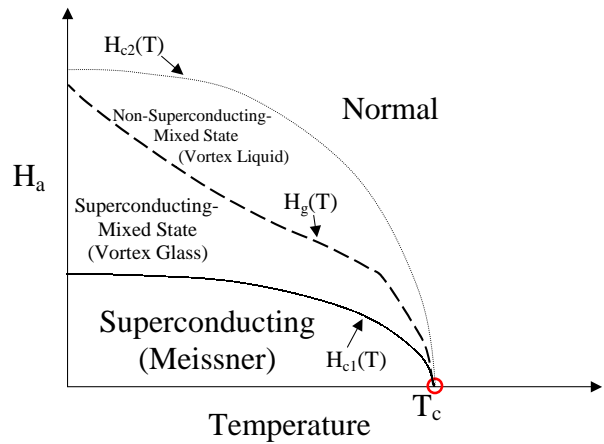

Fig. 2: Phase diagram for a type-II superconductor containing vortex liquid and vortex glass states.

In spite of the strong consensus that a phase transition to a state of zero resistivity should occur, we believe that the experimental evidence points toward the opposite conclusion.[13] In this paper we first discuss the application of scaling ideas to the analysis of current-voltage (I-V) characteristics. This powerful technique[11] is the basis for most of the analysis in the literature, including our own. We then use scaling to analyze our experimental data, and show that a serious problem arises: standard I-V scaling proves to be too flexible to allow definite conclusions to be drawn about the presence or absence of a phase transition. Finally, we discuss a signature of the phase transition that is required by scaling. This signature is not present in our data, or in the data in the literature that we have analyzed.

## 2. SCALING OF CURRENT-VOLTAGE CHARACTERISTICS

At first sight, answering the question "Is it superconducting?" seems straightforward. One need only attach leads to the sample, apply a current, and see if the voltage is zero or not. Of course, no voltmeter can say whether the voltage is strictly zero, but it is tempting to think that a large drop in the voltage, down to the limits of resolution of a good voltmeter, over a small (decreasing) temperature interval is an excellent indication that the sample has become superconducting.

Unfortunately, it is not this simple. In many instances, both theory and experiment agree that the resistance becomes very small very rapidly as the temperature is reduced, but does not ever become strictly zero. This occurs, for example, in superconductors in one dimension (D=1).[1] This reflects the fact that one-dimensional superconductors do not become fully superconducting: Thermal fluctuations destroy superconducting order for temperature T>0 when D=1. This means that, in a one-dimensional sample, the resistance will drop rapidly below the resolution of any voltmeter without ever becoming strictly zero.

The application of scaling ideas to superconductivity provides a powerful way around this problem. Scaling tells us that if a phase transition to a superconducting state occurs then the I-V characteristics are severely constrained. Thus, if I-V characteristics unambiguously obey scaling over a wide range of currents and (non-zero) voltages, the I-V characteristics provide very strong evidence for the phase transition. Failure to obey scaling would be strong evidence against the phase transition.

Fisher, Fisher, and Huse[11] have discussed scaling in superconductors in great depth. Here we provide heuristic arguments that lead to the scaling functions used to analyze experiments. We start with a key pair of physical quantities, the correlation length $\xi$, and the correlation time $\tau$.[14] The correlation length diverges at $T_c$, varying as

$$\xi \sim \varepsilon^{-\nu} \qquad (1)$$

where $\varepsilon = |T-T_c|/T_c$ and $\xi$ is the size of a typical fluctuation. Above $T_c$, $\xi$ is the size of superconducting regions that occur as fluctuations in the normal background, and below $T_c$, $\xi$ is the size of normal fluctuations in the superconducting background. The correlation time varies as

$$\tau \sim \xi^z \sim \varepsilon^{-z\nu} \qquad (2)$$

and is a measure of the lifetime of fluctuations. Eqs. (1) and (2) contain critical exponents $\nu$ and $z$ which characterize the diverging length and time scales.

The scaling hypothesis states that if a system is close enough to $T_c$, the singular behaviors of measured quantities (such as I, V, or the heat capacity, for example) are determined solely by $\xi$ and $\tau$. This is a very strong statement, and not an obvious one. The idea is that, as the length and time scales of fluctuations diverge, they eventually dominate the behavior of any measured quantity. Given the enormous success of the scaling hypothesis in explaining static and dynamic critical phenomena,[14] it seems safe to apply it to superconductors. To do this, dimensional arguments are constructed relating physical quantities of interest to lengths and times, thus determining their temperature dependences.

Since we are measuring I-V characteristics, we search for the scaling behavior of the microscopic counterparts of I and V, the current density J and the electric field E. We start with an argument for E. In a superconductor, flux is quantized in units of the flux quantum, $\Phi_o = h/2e$, where h is Planck's constant and 2e is the charge of a Cooper pair. Magnetic fields relevant to the transition scale with $\Phi_o$; for example, the upper critical field $H_{c2} = \Phi_o/2\pi\xi^2$ in mean-field theory.[1] This suggests that other quantities with units of magnetic field should scale as

$$B \sim \Phi_o \xi^{-2} \sim \xi^{-2} \qquad . \qquad (3)$$

By Faraday's law, $\nabla \times E = -\partial B/\partial t$. Using dimensional analysis inspired by scaling, this suggests

$$E/\xi \sim B/\tau \qquad . \qquad (4)$$

Combining Eqs. (2), (3), and (4) leads to the desired scaling form for E,

$$E \sim \xi^{-1-z} \qquad . \qquad (5)$$

Analogous to Eq. (3), Eq. (5) predicts how quantities with units of electric field should scale. We can use Eq. (5) and an energy argument to obtain the scaling form for the current density J. We equate the power dissipated in a fluctuation, $JE\xi^D$, to the thermal energy available over the lifetime of a fluctuation, $kT/\tau$, and solve for J to obtain

$$J \sim \frac{kT}{E\xi^D \tau} \sim T\xi^{1-D} \qquad (6)$$

where we have used Eqs. (2) and (5) in the final form. Note that the current density here is the D-dimensional current density; i.e., it has units of current/(length)$^{D-1}$. Note that Eq. (6) predicts how the critical current approaches zero for $T \leq T_c$, if the scaling hypothesis is correct.

The equations above can be used to calculate the scaling behavior of response functions, that is, they can answer questions like "How do fluctuations affect the conductivity above $T_c$?" In the limit of small current for $T>T_c$, we can use Eqs. (5) and (6) to calculate the fluctuation contribution to the conductivity, $\sigma'=J/E$, to obtain,

$$\sigma' \sim T\xi^{2+z-D} \sim T\varepsilon^{\nu(D-2-z)} \qquad . \qquad (7)$$

This fluctuation conductivity is added to the normal-state conductivity. The scaling hypothesis thus predicts how the conductivity diverges as $T_c$ is approached from above in terms of D, $\nu$, and z.

Eq. (7) can be extended to include nonlinear effects, such as those that occur at higher currents above $T_c$. Below $T_c$, where the current-voltage characteristics are not linear at small current, these non-linear effects are the leading-order effects. We write

$$\frac{E}{J} = \frac{\xi^{D-2-z}}{T} F_+ \left( \frac{J\xi^{D-1}}{T} \right) \qquad T \geq T_c \; ; \qquad (8a)$$

$$\frac{E}{J} = \frac{\xi^{D-2-z}}{T} F_- \left( \frac{J\xi^{D-1}}{T} \right) \qquad T \leq T_c \qquad . \qquad (8b)$$

When the argument of the unknown function $F_+$ is small, the function can be replaced by $F_+(0)$ which is a constant, thus Eq. (8a) is equivalent to Eq. (7) in this limit. From Eq. (6), it can be seen that the argument of the function is chosen to obey the scaling hypothesis: If Eq. (6) is true, varying J and T such that the argument of $F_\pm$ is constant does not change the physics of the superconductor. Eqs. (8) should thus be true over a wide range of currents and temperatures. The subscript $\pm$ on $F_\pm$ indicates that there are two functions, $F_+$ for $T \geq T_c$, and $F_-$ for $T \leq T_c$. Different functions are necessary to reflect the different physics above and below $T_c$, such as the fact that the low-current resistivity is zero below $T_c$, and non-zero above $T_c$.

It is convenient to have a scaling equation in terms of the directly measured quantities I and V. Using the fact that $V \propto E$ and $I \propto J$ (with the proportionality constants dependant on the sample shape), we write

$$\frac{V}{I} = \xi^{D-2-z} \chi_\pm \left( \frac{I\xi^{D-1}}{T} \right) \qquad . \qquad (9)$$

We have introduced two new unknown functions $\chi_\pm$ in Eq. (9), and have also dropped the first factor of $1/T$.[16]

It is instructive to take various limits of Eq. (9). For example, in the limit where $I\xi^{D-1}/T$ is small, the function $\chi_+$ approaches a constant, and the resistance in the limit of small currents for $T \geq T_c$ becomes

$$\frac{V}{I} \sim \xi^{D-2-z} \sim \varepsilon^{\nu(2+z-D)} \quad . \tag{10}$$

Another interesting consequence of Eq. (9) is obtained by considering the opposite limit, where T approaches $T_c$, the argument of $\chi_\pm$ approaches infinity, and the prefactor approaches either zero of infinity, depending on the value of D-2-z. For any non-zero current, we expect a non-zero voltage at $T=T_c$, since the critical current goes to zero at $T_c$. This will be true only if the factors of $\xi$ cancel in Eq. (9), or

$$\frac{V}{I} = \xi^{D-2-z}\chi_\pm\left(\frac{I\xi^{D-1}}{T}\right) \sim \xi^{D-2-z}\left(\frac{I\xi^{D-1}}{T}\right)^{(2+z-D)/(D-1)} \sim I^{(2+z-D)/(D-1)} \tag{11}$$

or

$$V \sim I^{(z+1)/(D-1)} \tag{12}$$

when $T=T_c$.

### 3. STANDARD SCALING ANALYSIS OF EXPERIMENTAL DATA

We next discuss the standard way in which Eqs. (9)-(12) are used to analyze data. We use data from a typical high-quality $YBa_2Cu_3O_{7-\delta}$ film. This films was laser ablated onto a $SrTiO_3$ substrate and had a thickness of 220 nm. X-ray diffraction indicated that the film was predominantly c-axis oriented. The film had a zero-field $T_c$=91.5 K and a transition width (10% to 90%) of 0.65 K. Resistance as a function of temperature in zero field is shown in Fig. (3).

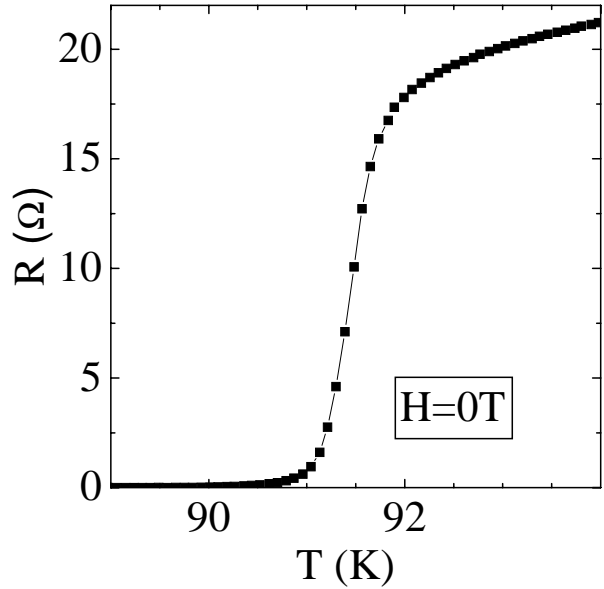

Fig. 3: Resistive transition in zero magnetic field for a $YBa_2Cu_3O_{7-\delta}$ film.

I-V curves for this sample were taken at various temperatures with a field of 4 T applied perpendicular to the film. These curves are shown on a log-log plot in Fig. 4 for temperatures between 70 K and 88 K.

A number of qualitative features are worth noting in Fig. 4. The I-V data at T=88 K are seen to fall on a straight line with slope 1. Since Fig. 4 is a log-log plot, a slope of 1 implies that V∝I, or that the I-V curve is ohmic over the entire range of current. This indicates that T=88 K is well above any superconducting transition temperature in a field of 4 T.

At somewhat lower temperatures, the data become non-ohmic (with steeper slopes on the log-log plot) at high currents, but always bend back towards ohmic behavior at lower currents. These ohmic "tails" have lower resistances at lower temperatures. The tails are the result of superconducting fluctuations, which reduce the resistance below the normal state value. Increasing the current suppresses the superconducting fluctuations, causing the steeper increase in voltage at higher currents.

As the temperature is lowered, the ohmic tails occur at lower and lower voltages and currents. At some point the tails disappear entirely, at around 81 K in Fig. 4. There are two possible explanations for this. It may be the temperature where a superconducting transition occurs in a field of 4 T, i.e. $T_g$=81 K when $\mu_0 H$=4 T. (We use the symbol $T_g$ to indicate the glass transition temperature. As discussed above, this could be a vortex-glass transition, a Bose-glass transition, or some other transition. The key point is that some type of transition that obeys scaling may occur at $T_g$.) A second possibility is that the ohmic tails persist to still lower temperatures, but are below the resolution of the voltmeter.

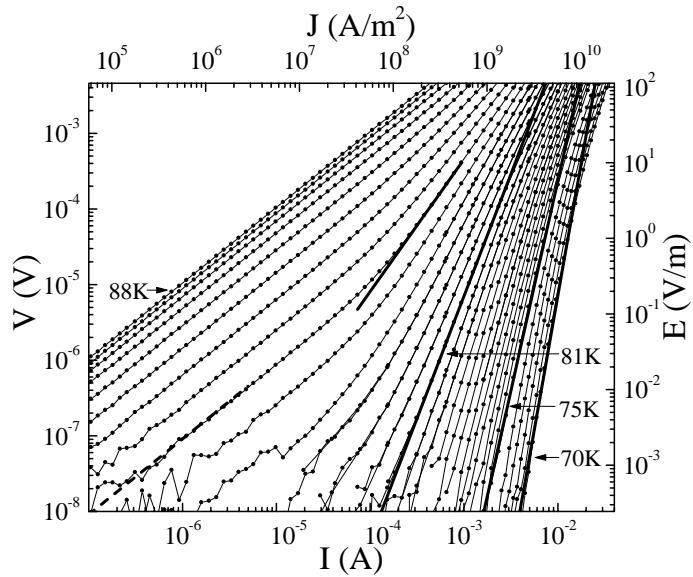

Fig. 4: I-V curves taken at constant temperature on a log-log plot. Dashed line has a slope of 1; solid lines are discussed in the text.

The standard analysis assumes that a transition does occur, and thus $T_g$=81 K. Assuming this to be correct for the time being, Eq. (12) predicts that the *critical isotherm* should be a straight line on a log-log plot, with slope given by $(z+1)/(D+1)$. The dark solid line drawn in Fig. 4 is a power-law fit to the critical isotherm. Using D=3 and Eq. (12), this determines a value of z=5.46. This value of z is consistent with those reported in the literature.[3-11]

Following the standard analysis, we next use $T_g$=81 K, z=5.46, and Eq. (10) to determine ν. The resistances $R_L$ are read off of the low-current tails in Fig. 4, and plotted on a log-log plot, as shown in the inset to Fig. 5 (a). It is seen that below about 87 K, a good fit is obtained, with deviations at higher temperatures. When Eq. (10) is fit within this temperature interval, the slope yields a value ν=1.5, again consistent with other values in the literature.[3-11]

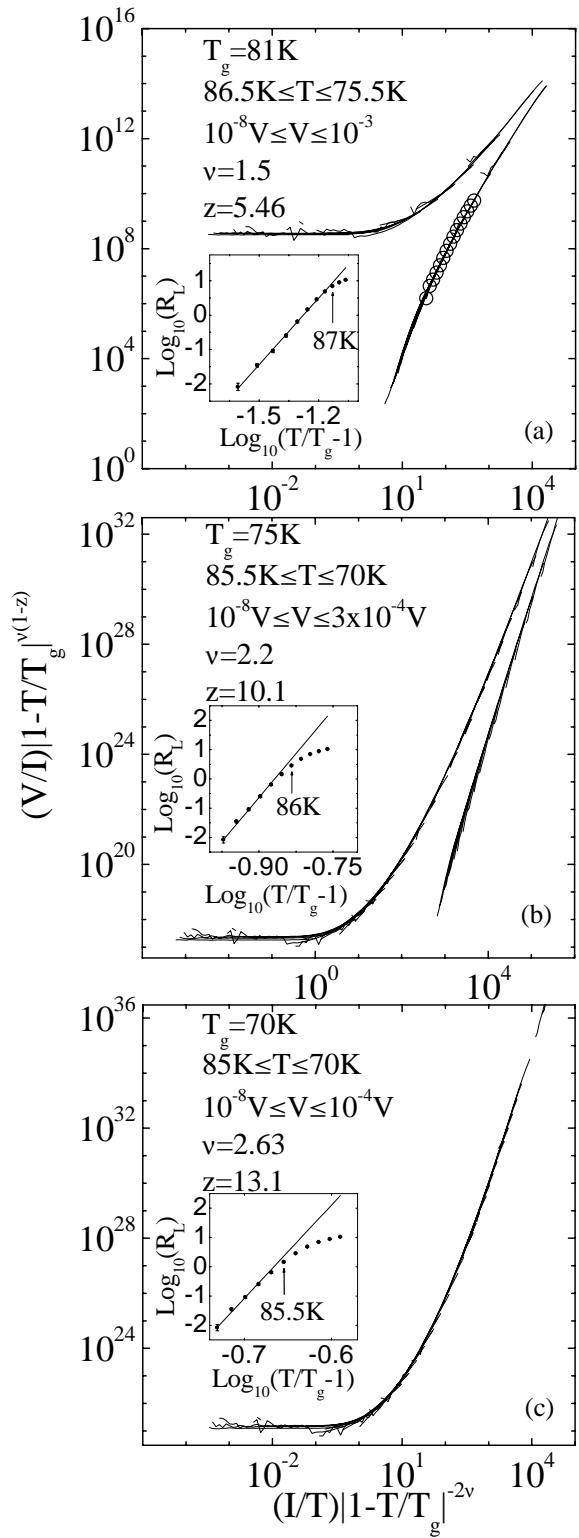

Fig. 5: (a) shows the data collapse resulting from the standard scaling analysis, with $T_g$=81 K. (b) and (c) show that good data collapse is obtained for other values of $T_g$, including the lowest temperature for which data were obtained.

The scaling equations are only expected to apply in a critical region close to $T_g$. We use the inset in Fig. 5 (a) to estimate the extent of the critical region, which is within ±5.5 K of $T_g$. In addition, data at high currents are not expected to obey scaling because the system is being driven too far from thermal equilibrium. An upper cutoff is conventionally set to the voltage where the critical isotherm begins to deviate towards ohmic behavior, which is at about $10^{-3}$ V in Fig. 4.

With $T_g$, z, and ν determined, we next re-write Eq. (9) as

$$\frac{V}{I}\xi^{2+z-D} = \chi_{\pm}\left(\frac{I\xi^{D-1}}{T}\right) \quad . \tag{13}$$

Eq. (13) predicts that a plot of $V\xi^{2+z-D}/I$ against $I\xi^{D-1}/T$ should "collapse" all of the data in the critical regime onto one of two curves, $\chi_+$ for $T \geq T_g$ and $\chi_-$ for $T \leq T_g$. This data collapse is shown in Fig. 5(a).

The data collapse shown in Fig. 5 (a) is very impressive to the eye. The apparent success of the data collapse is widely taken to indicate that the data scale. This, in turn, would indicate a phase transition has taken place. We show in the next section, however, that there are serious problems with this analysis.

## 4. A MORE CRITICAL ANALYSIS OF DATA COLLAPSE

While the data collapse in Fig. 5 (a) seems to be quite good, it does not eliminate the possibility that ohmic tails persist below the tentatively chosen value of $T_g$=81 K. We examine that possibility in this section.

First, we note that qualitatively, at least, *all* the isotherms with T≤81 K appear to be straight over some range in V in Fig. 4. They would thus *all* appear to satisfy Eq. (12), which suggests that $T_g$ may not be uniquely determined by the standard procedure. To test this idea, we re-did the standard scaling analysis with a different value of $T_g$=75 K. The result of this scaling analysis is graphed in Fig. 5 (b). Remarkably, the data collapse is also very good.

Taking this to the extreme case, Fig. 5 (c) shows the result of choosing $T_g$=70 K, the lowest temperature measured in the experiment. Here, since all the data are from temperatures above the nominal $T_g$, all the data collapse onto only one curve, corresponding to $\chi_+$ in Eq. (9). Once again, the collapse appears to be quite good.

It is alarming that seemingly good data collapse is obtained for any $T_g$ less than 81 K. One possible resolution of the problem is that one of the choices for $T_g$ gives better fits than the others. Unfortunately, it is not clear how to quantitatively determine whether fits to Eq. (13) are good or not, since Eq. (13) contains unknown functions. In the standard scaling analysis used on superconductors, data which is "too far away from the critical point" in temperature or current is discarded and parameters are varied until a "good" fit is obtained, with "good" usually being determined by eye.

We have proposed another approach designed to make the evaluation less subjective.[17,13] This approach is based on the fact that the unknown scaling functions have known limiting forms. At $T=T_g$ for small current, V should be a power of I, as can be seen from Eq. (12). For $T>T_g$, V should be linear in I for small currents, following a temperature dependence given by Eq. (10).[18]

Eq. (12) predicts that the critical I-V isotherm should be a straight line on a log-log plot. At least by eye, all the isotherms for T≤81 K in Fig. 4 seem to be straight. To clarify any differences between the I-V curves, we have plotted dlog(V)/ dlog(I) against log (I) in Fig. 6.

It is important to note that Eq. (12) predicts that the I-V characteristic taken at $T=T_g$ should be a horizontal line in Fig. 6. *None* of curves in Fig. 6 are horizontal lines, including the curve at T=81 K. While the solid line drawn over the T=81 K data in Fig. 4 appears to fit the data well, Fig. 6 makes it clear that deviations from the power-law behavior of Eq. (12) are systematic. This indicates that one of the key predictions of scaling is not obeyed by this data at any temperature.

Over most of the temperature range the curves shown in Fig. 6 are, in fact, qualitatively similar. Starting at the highest current, dlog(V)/dlog(I) first increases as the current is lowered, then reaches a maximum, then decreases. At the highest temperatures it is clear that the low-current value of dlog(V)/dlog(I) is one, i.e., the low-current behavior is ohmic. As the temperature is lowered somewhat, the low-current ohmic behavior is not seen, but the decrease in dlog(V)/dlog(I) is seen at lower currents. This is most probably due to the fact that the voltmeter has limited sensitivity. At still lower currents, even the maximum is not visible, again presumably due to limited sensitivity. The key point here is that once limited experimental resolution is taken into account all the curves shown in Fig. 6 are qualitatively the same. There is no clear indication for a transition at some specific temperature $T_g$.

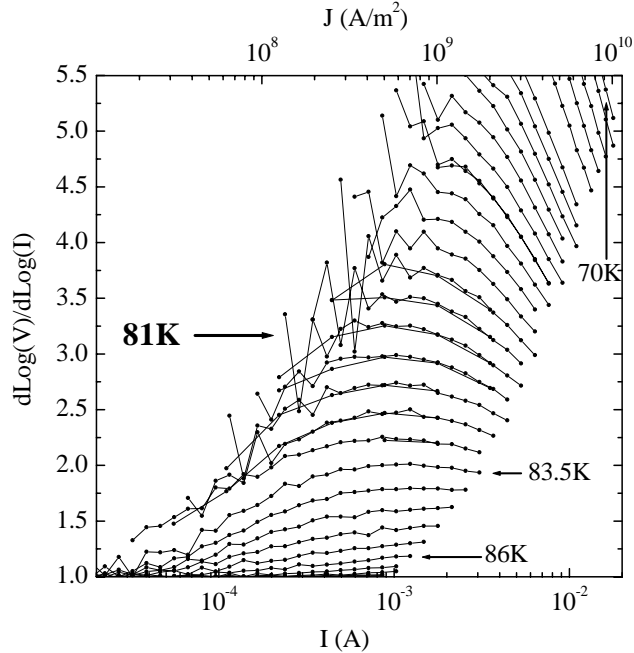

Fig. 6: Logarithmic derivative of the I-V curves plotted in Fig. 4. Note that, with the exception of a cutoff imposed by the voltmeter's sensitivity, all the curves are qualitatively similar: There is no qualitative change in behavior at the nominal $T_g=81$ K.

Derivate plots can be used in another way to test whether or not data scale. We note that Eq. (9) allows extrapolation of data to lower currents and voltages than can be measured. For example, Fig. 5 (a) shows the data points from the T=79 K I-V curve as open circles; the rest of the points on the lower curve in Fig. 5 (a) come from different temperatures. If scaling is assumed to work, however, we can extrapolate in the following way: We choose a temperature and a current, and assume the values $\nu=1.5$ and $T_g=81$ K, as is appropriate if Fig. 5 (a) is taken to represent a good data collapse. This determines a point on the horizontal axis of Fig. 5 (a), since this axis is given by $(I/T)|1-(T/T_g)|^{-2\nu}$. This determines a point on the scaling function, which in turn determines a point on the vertical axis. By setting this value equal to $(V/I)|1-(T/T_g)|^{-2\nu}$, we can solve for an extrapolated V using the parameters above and the value $z=5.46$. The results of such an extrapolation are shown in Fig. 7.

The downturn of the extrapolated curves for $T>T_g$, and the upturn for the extrapolated curves for $T<T_g$, have simple physical interpretations. They reflect the fact that in order to obey scaling, the low-current data must become ohmic at high temperatures, and they must become superconducting at low temperatures. If this behavior were seen in the actual data, it would provide strong evidence that a phase transition to a superconducting state is actually occurring. *The absence of this behavior in the actual data is strong evidence that a phase transition is not occurring.* The resistance is becoming very small very rapidly as the temperature is lowered, but there is no evidence that it is strictly going to zero at some non-zero temperature.

The change in sign of the slope of the extrapolated data in Fig. 7 corresponds to a change in concavity in Fig. 4. This led us to propose[13] an opposite concavity criterion as a test for the presence of a phase transition: At equal temperatures away from $T_g$ (i.e., temperature which have equal values of $|T-T_g|$), isotherms such as those in Fig. 4 should have opposite concavities at the same current if a phase transition has occurred. This criterion is not satisfied by our data, and is not, to our knowledge, satisfied by any data in the literature, indicating that a phase transition to a state of zero resistance does not occur in a magnetic field.

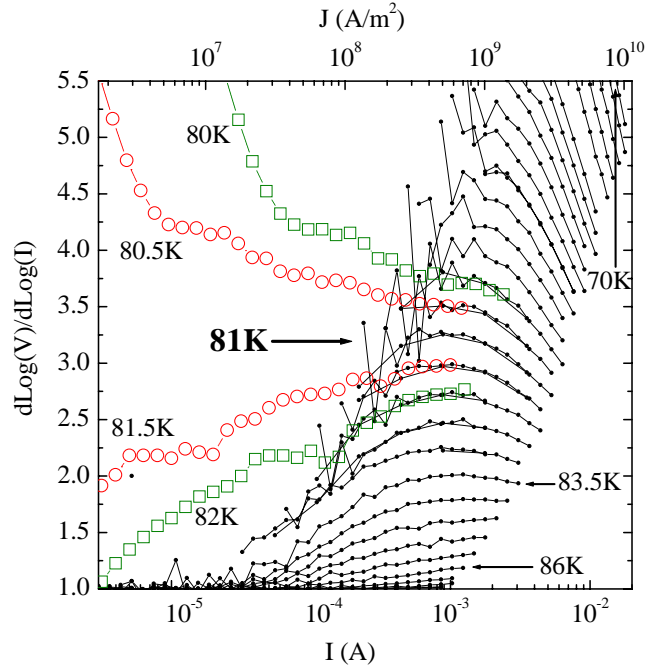

Fig. 7: Derivative plot similar to Fig. 6, with actual data shown as small solid points, and extrapolated data shown as open symbols. Note that the extrapolated data have a property not seen in the actual data: Curves with temperature below the nominal $T_g$ (T<81K) have a negative slope at low currents, while curves with $T>T_g$ have positive slope at low currents.

## 5. DERIVATIVE SCALING

The previous section showed that a derivative plot, Fig. 6, reveals systematic deviations from scaling predictions that are not obvious in the raw data plot, Fig. 4. Fig. 5 shows one problem with the scaling approach—it is clearly too flexible because it seems to allow any $T_g \leq 81$ K. Another problem that is not easily seen in Fig. 5 is that deviations from the scaling curve are systematic. In the following discussion we will show that scaling theory makes predictions about logarithmic derivatives that are very clearly not in agreement with experiment. While deviations from scaling of I-V data are not immediately evident in Fig. 5, they are clearly evident when logarithmic derivatives are scaled.

Starting with Eq. (9), we can evaluate dlog(V)/dlog(I) to obtain

$$\frac{d\log(V)}{d\log(I)} = \frac{I\xi^{D-1}}{T} \frac{\chi'_\pm\left(\frac{I\xi^{D-1}}{T}\right)}{\chi_\pm\left(\frac{I\xi^{D-1}}{T}\right)} \equiv \Lambda_\pm\left(\frac{I\xi^{D-1}}{T}\right) \quad . \tag{14}$$

We thus see that dlog(V)/dlog(I) should only depend upon $I\xi^{D-1}/T$ if data obey scaling. In a manner analogous to Eq. (13), Eq. (14) predicts that dlog(V)/dlog(I) plotted against $I\xi^{D-1}/T$ should cause all the data to fall onto one of two curves, $\Lambda_+$ or $\Lambda_-$, depending on whether $T>T_g$ or $T<T_g$.

Fig. 8 is a plot made to test Eq. (14). Although the data for $T>T_g$ and $T<T_g$ fall into two different regions of the plot, it is clear that the data do not scale. Such systematic deviations can be seen in expanded versions of Fig. 5, but they are much more visible in Fig. 8. This is further strong evidence that a phase transition is not occurring.

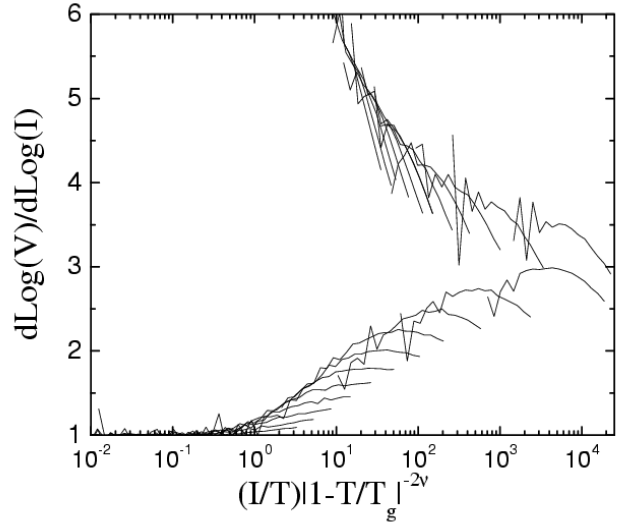

Fig. 8: Plot of dlog(V)/dlog(I) against $I\xi^{D-1}/T$. According to Eq. (14), all of the data should fall onto one of two curves, depending on whether $T>T_g$ or $T<T_g$. The predicted data collapse does not occur.

## 6. DISCUSSION AND CONCLUSIONS

Based upon the apparent success of scaling, a consensus has emerged that a superconducting phase transition occurs in a magnetic field. There have been, however, papers which questioned this consensus.

One approach is to produce simulated I-V curves based on models without phase transitions. When this was done, it was shown that the simulations could be collapsed onto scaling functions,[19,20] demonstrating that agreement with scaling alone does not imply a transition. It was also argued[21] that experimental data on $YBa_2Cu_3O_{7-\delta}$ can be fit to both scaling functions (which would suggest a phase transition) and to thermally-activated flux motion (which would suggest that a phase transition does not occur), again showing that conclusions based on apparent agreement with scaling are not iron-clad.

Further serious theoretical concerns have been raised. While early theoretical work in support of the vortex-glass picture (in the presence of point disorder) neglected the effects of screening, it was found that the inclusion of screening destroyed the transition in D=3.[22] Monte-Carlo simulations have supported this scenario.[23] Finally, recent careful simulations of vortices undergoing random thermal motion showed "window-glass" dynamics; again, no true phase transition was observed.[24]

Our work, reported in this paper and in Ref. 13, is in agreement with the absence of a phase transition view. Derivative plots such as Fig. 6 provide strong evidence against a phase transition, especially when compared with extrapolations which assume a transition, Fig. 7. Further strong evidence is provided by the fact that the derivative data do not obey scaling, and that the disagreement is very clear (Fig. 8). We have also shown that the standard I-V scaling approach is far too flexible, as shown in Fig. 5. In particular, *any* choice of $T_g$ below the standard choice gives data collapse that seems to look good. If a phase transition does not occur in a magnetic field, then the conventional choice of $T_g$—the first isotherm without a visible ohmic tail—would depend upon the shape of the sample and the resolution of the voltmeter. This last point may explain the wide range of critical exponents reported in experiments.

Further work is clearly needed, but there is much new evidence that superconductors in a magnetic field do not undergo a phase transition to a state with zero resistance.

## ACKNOWLEDGEMENTS

We would like to thank Alan Dorsey for many useful discussions of vortices in superconductors, and especially for his explanation of scaling, which was the basis for Section 2 of this paper. We are also grateful to Steven Anlage for many useful discussions of experiments and scaling (both AC and DC), as well as Thomas Frederiksen for his insightful comments and questions and his critical reading of this manuscript. This work was supported by the National Science Foundation through Grant No. DMR-9732800 and the Maryland Center for Superconductivity Research.